PAPER

# Scalable CI/CD for Legacy Modernization: An Industrial Experience Addressing Internal Challenges Related to the 2025 Japan Cliff

Kuniaki Kudo[†], *Nonmember* and Sherine Devi[††], *Nonmember*

**SUMMARY** We have developed a Scalable CI/CD Pipeline to address internal challenges related to Japan 2025 cliff problem, a critical issue where the mass end of service life of legacy core IT systems threatens to significantly increase the maintenance cost and black box nature of these system also leads to difficult update moreover replace, which leads to lack of progress in Digital Transformation (DX). If not addressed, Japan could potentially lose up to 12 trillion yen per year after 2025, which is 3 times more than the cost in previous years. Asahi also faced the same internal challenges regarding legacy system, where manual maintenance workflows and limited QA environment have left critical systems outdated and difficult to update. Middleware and OS version have remained unchanged for years, leading to now its nearing end of service life which require huge maintenance cost and effort to continue its operation. To address this problem, we have developed and implemented a Scalable CI/CD Pipeline where isolated development environments can be created and deleted dynamically and is scalable as needed. This Scalable CI/CD Pipeline incorporate GitHub for source code control and branching, Jenkins for pipeline automation, Amazon Web Services for scalable environment, and Docker for environment containerization. This paper presents the design and architecture of the Scalable CI/CD Pipeline, with the implementation along with some use cases. Through Scalable CI/CD, developers can freely and safely test maintenance procedures and do experiments with new technology in their own environment, reducing maintenance cost and drive Digital Transformation (DX).
*key words: 2025 Japan Cliff, Scalable CI/CD, DevOps, Legacy IT Modernization.*

## 1. Introduction

In 2018, Japan's Ministry of Economy, Trade, and Industry (METI) published a report in which they identified a problem called the 2025 Cliff [1]. If left unaddressed, it is projected that Japan could lose up to 12 trillion yen per year after 2025, which is 3 times more than the cost in previous years. As comparison, Japan's defense spending budget is set at 8 trillion yen per year, therefore an issue that could potentially result in losses exceeding that amount should be regarded as a major challenge, one that will impact Japan's national interest and competitiveness in the future. At the core of this issue, there is an impeding mass end of service life (EOSL) of the legacy IT systems that is currently still utilized across numerous Japan enterprises [2]. Many of these systems were developed decades ago with legacy technology and inadequately documented code, making it increasingly challenging to maintain, update and replace. As the EOSL period approaches, the risks associated with system failures escalate, potentially leading to disruption in business operation.

Furthermore, the reliance on these legacy systems has imposed a significant burden on IT engineers, who now need to allocate substantial time and resources to maintenance and operational tasks, where more work and attention could have been done on innovation and development of new technological solutions. This hindered progress in Digital Transformation (DX) initiatives, creating technological gap which will leave us behind [3].

We at Asahi also faced the same internal challenges regarding legacy system that reflect the issues identified in the 2025 Cliff problem. Over the past decades, system changes and maintenance relied heavily on manual procedures, which can only be done by limited number of personnel in limited number of testing environment. These factors have made updates or changes seen as a high-risk task, leading to reluctance to do any update or modification to the system. Consequently, system maintenance was neglected, resulting in outdated middleware and operating systems that are now nearing their end of service life (EOSL) thus require a very high maintenance effort and cost. Further complication also arises from the "black box" nature of these legacy systems, where no single individual has the knowledge of the complete underlaying processes. Operations and maintenance are conducted based on previously written manuals, making updates and changes not only difficult to implement, but also challenging to verify. If these issues persist, maintenance costs will continue to increase significantly every several years, without any development of new functionalities to the system.

We have initiated our modernization effort by incorporating CI/CD concept into the current process to address these challenges. We have developed a Scalable CI/CD framework that enables parallel creation of multiple test environments that could be scaled automatically based on demand. This framework automates build and deployment of a system to an isolated environment, eliminating the needs of manual intervention in the process. By leveraging this approach, developers can safely and efficiently experiment with multiple code updates, changes, maintenance procedures each in their own separated and isolated environments. When the changes or procedures are verified to be working as expected, they can then implement it to the production environment. This will ultimately reduce the initial reluctance to implement updates, as developer no longer need to fear unintended consequences. Moreover, the time and effort required for maintenance can be significantly reduced, allowing developers to focus on innovating and developing new solutions.

---

[†]The author is with Asahi Group Holdings, Tokyo, 130-8602 Japan.

[††]The author is with Asahi Group Holdings, Tokyo, 130-8602 Japan

Through this industrial experience paper, we aim to share our efforts in reducing manual processes and modernizing legacy system process through a Scalable CI/CD framework. Our goal is to provide practical insights for organizations facing internal challenges that align with issues identified in the 2025 Japan Cliff problem.

## 2. Problem Statement

The current process at Asahi for updating and modifying system codebase highly relies on manual process. To illustrate this, consider the typical workflow in a development environment of a Java EE application for a particular system in Asahi as shown in Fig 1. When developers update a code, the changes need to be manually fetched from the code repository and built in a build environment using a development server. The resulting build(s), in this case a compressed Java enterprise archive folder (ear Folder) are then manually copied to a deployment environment, in this case using a file transfer client like WinSCP. Here the deployment scripts are executed manually to deploy the builds to an operating or QA environment. Finally, developers will check the operating environment to verify whether the code changes produce the expected results and execute test cases as necessary, which is typically done sequentially so it also takes time and effort. This manual process is not only labor-intensive but also prone to human error, increasing the risk of mistakes during build or deployment stages.

As seen from the workflow, current process also relies on multiple dedicated servers for different stages of the workflow that needs to be continuously maintained. But even though there are several servers utilized as the environment for different stages, but there is only limited number of QA environment that is available to verify the changes or updates in. This limitation restricts developers from experimenting with multiple changes simultaneously. This combination of manual process and limited number of QA environment results in psychological barrier for developers from doing updates or changes, since developers are afraid that even minor changes could disrupt the currently stable processes.

Based on these problems, a solution is needed to remove the psychological barrier and make developer feel safe to try and experiments with any changes or updates as they want. To achieve this, we should create a framework where each developer can have their own dedicated QA environment, it is separated and isolated from each other so their changes will not affect the current running process, hence they can develop and experiments with the code freely and safely. This framework should also automate the entire build, deployment, and testing process. It should also eliminate the need for maintaining unused servers and could provide numerous QA environments that can be scaled automatically, meaning it can be dynamically created and deleted as needed. It should also enable parallel automated test which could shorten the duration of testing. Additionally, incorporating mechanisms for real-time error notifications and testing reports will also be helpful, ensuring the developers and stakeholders are promptly informed of build failures and can verify the results of the changes or updates don't break the current functionality before it is merged.

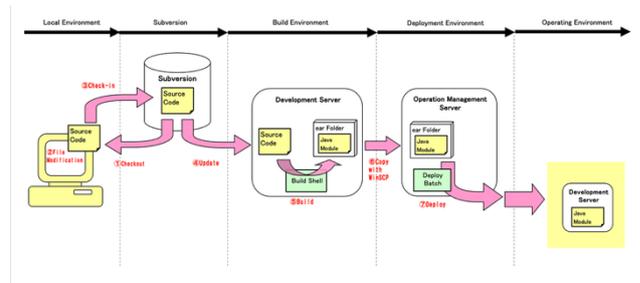

Fig 1 Existing manual workflow

## 3. Related Works

Previous works have examined the implementation of CI/CD pipelines to improve software development process by reducing manual effort, improve software quality, and enabling faster delivery.

Naveen et al. [4] explored the impact of building CI/CD pipelines with Jenkins automation on improving web application development, it highlights some benefits including early issue detection, increased developer focus, time savings, and enhanced reliability. Dileepkumar and Mathew [5] presented an overview of Jenkins based CI/CD best practices in achieving reliable and efficient CI/CD pipeline, which results in quicker delivery time, higher quality software, and greater team cooperation.

Some previous works also utilize containerization technologies such as Docker and cloud platforms to improve CI/CD workflows. Vernekar et al. [6] presented the integration of Docker and Jenkins in a CI/CD pipeline for a hotel reservation app, where throughout several phases of development, testing, and deployment, the integration of Docker guarantees a consistent and portable environment. Chatterjee and Mittal [7] presented a comparative analysis between manual and automated deployment strategies, highlighting improvements in product quality and customer satisfaction through CI/CD and DevOps integration using tools such as GitHub, Docker, and AWS EKS.

Other previous work focus on proposing architectural design for CI/CD pipeline. Patchkaew et al. [8] propose an Adaptive CI/CD architecture that divides the CI/CD framework into seven distinct layers from Source Code Management to Monitoring Tools, which would serve as a guideline that can adapt to future technology trends to enhance the quality of the software development process.

While these previous works have already highlighted the implementation and benefits of CI/CD with various technologies, most implementations assume a single shared QA environment or a fixed number of test environments, which may introduce limitations in larger scale enterprise settings where multiple developers need to validate various changes. In this paper, we share our experience extending the existing CI/CD practices by designing and implementing a Scalable CI/CD framework that allows automatic provisioning and deprovisioning of isolated QA environments on demand, which will enable developers to independently test and verify changes without interfering other ongoing processes, hence supporting safer and faster development workflows.

## 4. Design and Architecture

Our Scalable CI/CD pipeline is designed to address the challenges with the current system that is outlined in the problem statement. This section provides an overview of the system components requirements, design, the key components, and the pipeline architecture of this Scalable CI/CD Framework.

4.1 System Components Requirements

The Scalable CI/CD framework requires several components that satisfies the following conditions:
- CI/CD orchestration tool: a CI/CD orchestration tool that can receive webhook events, manage CI/CD pipelines per branch, manage nodes that could execute commands and tasks on cloud platform.
- Source code management system: a Git-based repository that supports branching and can send webhook events (branch creation, update, or deletion, and pull request opened or closed events).
- Cloud computing platform with container orchestration: A platform that supports dynamic creation and deletion of serverless containerized environments, and the ability to launch containers, manage container images, and provide shared storage for artifacts.

For our implementation, we chose Jenkins as CI/CD orchestration tool, GitHub for source code management system, and AWS as cloud computing platform because it meets these requirements. However, the proposed method itself is not dependent on these specific tools and could be applied using alternative technologies that provide equivalent capabilities.

4.2 Design Overview

Fig 2 illustrates our proposed design overview of the Scalable CI/CD Framework, where the build and deployment environment are refactored to serverless containerized environment. The Scalable CI/CD automates the entire build, deployment, and testing process using Jenkins Pipeline. GitHub is used for code repository, and GitHub Webhooks is utilized to trigger a Jenkins Pipeline to start automatically whenever there is a code changes in the repository. The Jenkins pipeline will then orchestrate an end-to-end CI/CD process. Starting with the build phase, AWS Fargate is utilized to provision serverless containerized build environment where the source code is built, and the environment will be automatically deleted upon completion of the build. The build result is then directly deployed to a deployment environment, which is also serverless and containerized through AWS Fargate. After deployment, testing phase is done where test cases could be run parallelly, which for each test case a dedicated Fargate task is launched and then execute a test script. After all the test cases are executed, test reports are created visualized in the Jenkins dashboard. Each branch in the code repository will run its own CI/CD pipeline and have their own corresponding environment isolated from each other. The environment will also be automatically deleted if the branch is deleted.

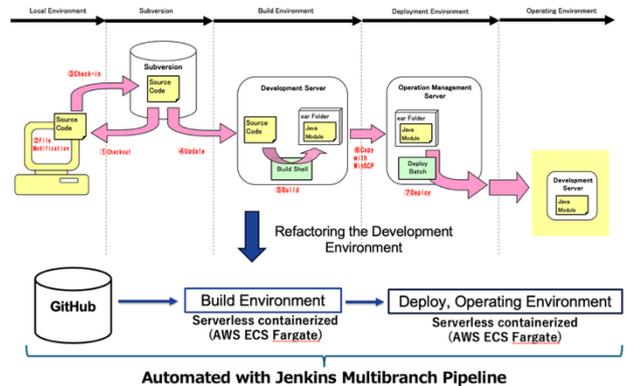
Fig 2. Proposed Design Overview

If there is error in the process, an automated error notification system will send an error alert to a designated target, in our case, to a Slack communication channel. This ensures rapid response and resolution of issues, as developers and stakeholders are always informed on the current progress.

4.3 Framework Components

The Scalable CI/CD framework is built using several key components and services. Below we detail the components and their role in the overall framework.

4.3.1 Jenkins

Jenkins is an open-source automation server [9] used to orchestrate the CI/CD workflow. Our Main Scalable CI/CD pipeline is implemented using Jenkins Multibranch Pipeline Job, which automatically detects and processes individual branches in the GitHub repository. This allows each branch to have their own isolated CI/CD pipeline running, which enables it to set up different isolated development environment for each branch.

4.3.2 GitHub

GitHub is a cloud-based platform where we can store, share, and work together with others to write code [10]. GitHub is used as the source code management system (SCM) for the Scalable CI/CD Pipeline. We use trunk-based development, where each new feature or fix is developed in separate branches before it is merged to the main branch. Branches in trunk-based development are typically short-lived and are merged frequently into the main branch. This approach aligns with Scalable CI/CD, where quick verification of changes on each branch is possible because each branches has its own QA environment, and as a result, feature or fixes branches can be merged into main branch quickly.

4.3.3 Amazon Web Services (AWS)

Amazon Web Services (AWS) [11] is a cloud computing platform which provides the infrastructure for running and scaling the CI/CD Framework.
The Scalable CI/CD Framework leverages several AWS services:
· EC2: virtual machine instances used to host Jenkins server and nodes for executing the CI/CD pipeline.
· Elastic Container Registry (ECR): container registry where container images for build, deployment, and testing environments are stored.
· Elastic File System (EFS): file storage service used as volume for containers.
· Elastic Container Service (ECS): main service that is used for deploying and running the containerized application. Within ECS, tasks are executed based on predefined task definitions, which serve as a blueprint for running containerized applications. The task definition is then launched as an ECS Fargate Task within a designated ECS cluster.

The framework utilizes ECS Fargate for its serverless capabilities, allowing dynamic and quick resource provisioning and deprovisioning as needed without the needs to directly manage the servers or infrastructure. This enables the Scalability feature of our Scalable CI/CD framework.

4.3.4 Docker Container

Docker [12] is a containerization platform that packages applications and their dependencies into containers. We created custom docker images for build, deployment, and test environment, and we pushed and stored it to AWS Elastic Container Registry (ECR). These images are then used when running ECS Fargate Task containers to create standardized reproducible environment.

4.4 Pipeline Workflow

The Scalable CI/CD Framework includes two distinct Jenkins Pipeline, each designed to handle specific events. These pipelines together automate the creation and deletion of isolated QA environments. Below, we describe the workflow of each pipeline.

The first or main pipeline is triggered when a new branch is created, or an existing branch is updated in the GitHub Repository, or a pull request is created. Fig 3 illustrates the pipeline workflow. This pipeline orchestrates the end-to-end CI/CD process, including build, deployment, and testing. The workflow consists of several steps. First step is initialization, when a developer pushes a new branch or updates an existing branch or creates a pull request, GitHub will send a webhook event to Jenkins, which will start a Jenkins Pipeline Job for that branch or pull request, utilizing the specified Jenkins Node. The branch's code is cloned into the Jenkins workspace within the Jenkins Node. In the case of a pull request, the cloned code corresponds to the projected merge result with the main branch A folder is created in the workspace and mounted to an Elastic File System (EFS) volume. This EFS folder is populated with the relevant source code, ensuring that code and artifacts are accessible across containerized environments. The pipeline then dynamically creates and registers ECS Task Definitions for the build, deployment, and testing processes. These task definitions specify the container images (stored in ECR), configurations, and mount points for the EFS volume. We specified the mount point of the containers volume as the mounted folder we created before. This ensures that the Fargate tasks launched later has access to the necessary code and artifacts.

Next is the build step, where a Fargate Task is launched using the task definition configured for creating build environment. This Fargate task will provision a containerized build environment using the container image and execute scripts to build the source code. Since the container volume are mounted to the EFS, the build results are stored in EFS. Upon build completion, the Fargate task is stopped automatically, and the pipeline proceed to the next step.

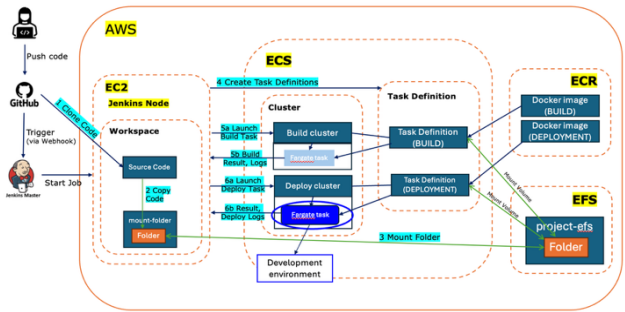

Fig 3 Pipeline Workflow

The next step is deployment step. First, if there is an existing deployment environment associated with the branch or pull request (from previous creation or updates), it will be stopped. Then, a new Fargate Task is launched using the deployment task definition. This task provisions a containerized deployment environment with the specified container image and mount volume. Since it is mounted to the same EFS folder, the Fargate task could access the build result. It will then deploy the build results to this environment. The environment is accessible via the Fargate task's IP address, allowing developers to verify the deployment results. This environment is persistent and remains available for each active branch or pull request, serving as the isolated QA environment where developers can verify the changes.

Optionally, a testing step can be integrated into the Jenkins Pipeline to automate the validation of the deployed application. Once the deployment process is confirmed to be successful, the pipeline retrieves the IP address of the deployment environment. For each previously created test case, a separate Fargate Task is launched to execute the test script against the deployment environment and outputs the result. These Fargate Tasks will then stop immediately upon test completion. This way, we can parallelize and automate testing process and thus reduce testing time. Since these Fargate tasks are mounted to the same EFS, test results are accessible in the Jenkins workspace. The pipeline then aggregates these results and generates the API Test report chart which is displayed in the Jenkins Portal.

The second pipeline (Sub Jenkins Job) ensures that the deployment environment associated with the deleted branch or closed pull request are promptly cleaned up and deleted. This pipeline is triggered when a branch is deleted, or a pull request is closed. When this happens, GitHub will send a webhook event to Jenkins, and the pipeline will start by verifying the GitHub event and getting the branch or pull request name. It will then find the running ECS Fargate Task (which is the QA environment) for this branch or pull request and then stop the task (delete the environment). Utilizing both Jenkins Pipelines, we can automate the creation and deletion of QA environment, thus making this CI/CD framework to be scalable.

**5. Implementation**

This section details the technical implementation of key components within our Scalable CI/CD Framework.

5.1 Target Applications Conditions

The proposed method is applicable to any application that can be built and deployed in a containerized environment. In our case, we have implemented Scalable CI/CD framework in web applications, including Web APIs and Front-end web applications, and backend enterprise application like Java EJB-based application. The proposed method does not impose specific requirements regarding the types of automated test. The type of tests to be included in the pipeline depends on the project's requirements and can be tailored by the team implementing the CI/CD framework. For example, a Web API project may include automated API tests step in the Jenkins pipeline that validate various API endpoints responses against expected response.

5.2 Container Images

Containerization sometimes involves packaging both the environment and application build artifacts within a single image, which mean each code modification requires image rebuild. Since we need to create a different environment for each branch, using this approach would mean multiple images rebuilding every time a new environment is launched. Hence our approach decouples the environment configuration from the application artifacts. The framework maintains base images that contain only the required environment configurations, stored in AWS Elastic Container Registry (ECR). This base image can then be used to create multiple containers, one for each branch, allowing each branch to have its own isolated environment. For instance, we have one base docker image for build environment, but for each branch, we will have separate containers spun up. Since the container volumes are mounted via AWS Elastic File System (EFS), we can inject the specific code for each branch into the corresponding container. This allows the code for each branch to be built independently within its isolated environment.

We created three container images for three distinct environments: build, deployment, and test. These images contain the necessary tools and dependencies for each environment. When containers are launched through Fargate Task, application specific artifacts such as source code, build result, or test cases script are dynamically mounted via AWS Elastic File System (EFS), eliminating the need for continuous image rebuilds. This separation enables efficient resource utilization while also maintaining environment consistency across pipelines.

5.3 Elastic File System (EFS)

As described section 4.1 we use AWS EFS as mount volume for the running containerized environments. To maintain isolation between concurrent pipeline runs, each branch will create and maintain its own dedicated directory within the EFS. This enables parallel execution of multiple pipeline instances without risk of overwrites between branches. We implement branch isolation through customized ECS task definitions, where container mount point path is dynamically assigned to branch specific EFS directories as they are created. This ensures that parallel operations on different branches can proceed with data integrity maintained.

5.5 Elastic Container Service (ECS)

We launched all the containerized serverless environments through AWS Elastic Container Service (ECS) Fargate Task. The implementation consists of creating task definitions and clusters. For task definitions, we specify some parameters which defines the container run such as container image URIs sourced from ECR, port mapping configurations, environment variable declarations, volume mount specifications linking the container volume to branch-specific EFS directories, and execution commands for the container as needed. For clusters, we created separate clusters with AWS Fargate (serverless) infrastructure, each for build, deployment, and test process to improve monitoring and tracking of Fargate tasks based on operation type.

5.5 Jenkins Pipeline

Jenkins pipeline is the backbone of the Scalable CI/CD framework. It automates the build, deployment, and testing process. This subsection details the implementation of the two Jenkins Pipelines this framework use: Main Jenkins Job and Sub Jenkins Job.

The Main Jenkins Job is implemented as a Jenkins Multibranch Pipeline, which is triggered when a branch is pushed or updated in the GitHub Repository. We chose Jenkins Multibranch Pipeline as it allows for each code branch to have their own pipeline running, ensuring that each branch is independently build, deployed, and tested. In the pipeline settings, we configure the branch sources as the GitHub repository we have created for the system source code, and we specify the path to the Jenkins script in the GitHub repository. We also configured the Jenkins Job to scan the repository by webhook and specify the trigger token to be used. Below, we outline the stages of our current Main Jenkins Job Pipeline as configured in the Jenkins script included in the codebase. These stages are customizable and could be adjusted further as needed.

- Checkout stage: pipeline begins by checking out the code from the branch in the GitHub Repository and copies it to the workspace in Jenkins Node.
- Mount volume stage: a folder is created in the workspace and mounted to an EFS. Relevant codes are then copied to this folder. This mounted folder will be utilized as the volume for the containers.
- Create ECS Task Definition stage: task definitions for build, deployment, and testing process are created and registered with ECS. The pipeline uses JSON-formatted templates that are already created previously, and only specific parts (mount volume name) are modified to the folder path in the previous stage.
- Start Build & Check Result stage: this stage begins by launching ECS Fargate tasks, each to build the frontend and backend code, and will run parallelly. The pipeline will wait for the build tasks to finish by periodically checking their status. Once finished, the pipeline will retrieve the build logs and verify the result. Successful builds proceed to the next stage.
- Deployment stage: this stage begins by stopping any existing ECS Fargate task associated with the branch that is currently running (from previous builds) by checking the branch name tag. A new ECS Fargate task is then launched using deployment container image, and the branch name is tagged to the task for identification. The build results are then deployed to this deployment environment.
- API Test stage: ECS Fargate task are launched for each API Test Case. The pipeline waits for all test case to complete and stores the XML file results of the test cases within the mounted EFS folder.
- Generate Report: using the JUnit plugin [13] in Jenkins, the pipeline will generate a detailed API Test Report from the XML file results of the test cases performed in previous stage.
- Clean up stage: final stage involves cleaning up the now unused resources. Here the mounted folder contents are deleted, EFS volume is unmounted, unused task definition are removed, and Jenkins workspace is cleared.

We use a try-catch-finally block to encapsulate these stages, so when an error occurs in any of the stages,

the pipeline will stop any existing Fargate Task for the branch, execute the clean-up stage, and send an error notification alert whose implementation will be shown in subsection 4.6.

The second Jenkins Job, Sub Jenkins Job is triggered when a branch is deleted, or a pull request is closed in a GitHub repository, which will then clean up the environment associated with the branch. A webhook will be sent to a Jenkins Job and it will stop the corresponding Fargate task of the branch and delete the QA environment, which prevents unnecessary costs and resource allocation.

The job is implemented as General Jenkins Pipeline, and is configured to use Generic Webhook Trigger, which listens for webhook events from GitHub. A variable is defined to capture the webhook payload, and the x-github-event header is added in the request to enable identifying the type of event that is sent from GitHub. Configure the token name to be used.

The pipeline script implements branch environment clean up through the following process. First, the event type is validated to identify branch deletion or pull request closure events. Next, the branch or pull request name is extracted from the webhook payload. It then identifies the current running ECS Fargate Tasks and stops any tasks associated with the branch by matching the branch name tag.

5.6 Automated Testing in Scalable CI/CD

As mentioned earlier, each branch has its own isolated deployment environment and multiple test case can be run in parallel against one environment. Here we will illustrate the concept with Web API testing. We created several API test cases script in python and built a base Docker image for the test environment, in this case a python image with requests and pytest libraries installed. An "API Test" stage is defined in the Jenkins pipeline, where here the Jenkins node retrieves the IP address of the deployment environment associated with the branch, and starts the automated testing process for that environment. The testing process utilizes the same concept of serverless containerized environment. Using the base test environment image, the Jenkins node spun up separate containers (launch Fargate tasks) for each test case. For example, if there are five test cases, five Fargate task will run in parallelly. Each Fargate task executes a different test case from the prepared test case scripts and generate XML file for the test result to be used in the next stages. For example, one of the test case might verifies the data retrieval endpoint of the Web API, so it sends a GET request to the deployment environment IP address to retrieve the data and asserts that the response status code is 200 and that the returned data matches the expected values, and then output a XML file for the test result. The pipeline waits for all test cases to complete before proceeding to the next stage.

5.7 GitHub Webhook

We configured two webhooks in the GitHub repository each designed to trigger different Jenkins pipeline based on the type of event. We added the webhook through the Add Webhook functionality in the Settings options of the GitHub repository. The first webhook is configured to trigger the Main Jenkins Pipeline whenever a push event occurs in the repository. For this webhook, we use multibranch webhook trigger [14] type. The payload URL for this webhook consisted of the Jenkins server URL, multibranch-webhook-trigger and a pre-configured token, for example: https://{your jenkins URL}/multibranch-webhook-trigger/invoke?token={your token}. The content type is set to application/json and the webhook is configured to be sent only when push events happen.

The second webhook is configured to trigger the Sub Jenkins Pipeline when a branch is deleted, or a pull request event happen. For this webhook, we use generic webhook trigger [15] type. The payload URL for this webhook consisted of the Jenkins server URL, generic-webhook-trigger and a pre-configured token, for example: https:// {your jenkins URL}/generic-webhook-trigger/invoke?token={your token}. The content type is set to application/json and the for the type of event, we chose individual events and selected branch or tag deletion and pull requests.

5.8 Error Notification Alert

To improve observability and support faster response to pipeline failures, we integrated Jenkins with Slack to send error notifications to a designated communication channel. This integration is achieved using the Slack Notification Plugin [16] for Jenkins. Below, we detail the implementation of this feature.

In Slack, a slack app is created using the template YAML file provided by the Jenkins Slack Notification Plugin [16]. The app includes a bot user that could send the error alert messages to Slack channels. The app is installed in the Slack workspace, and the necessary permissions are granted.

In Jenkins, the bot user token generated during the app creation process is stored as a secret text credential and used in the slack configuration of Jenkins system. A default Slack channel is specified, and custom slack app bot user option is enabled. Then the Jenkins bot user is invited to the Slack channel where the error notifications would be sent.

The error notification functionality is integrated into the Jenkins pipeline script using the slackSend pipeline step. In the catch error block of the Jenkins script, we added slackSend to send error notification whenever an error occurs during pipeline execution. The error notification could be customized with necessary details, in our case we included the repository name, branch name, Jenkins Job URL, and the build number.

5.9 Limitations and Constraints

The Scalable CI/CD framework assumes that the target applications can be built and run within containerized environments. Software for embedded system that is tightly coupled with specific hardware or requires direct access to physical resources may not be suitable for this approach. Additionally, the current implementation is developed based on AWS services (ECS with Fargate, ECR, and EFS). As such, while the framework is designed not to tie to specific cloud computing platform, portability to other cloud platforms will require additional adjustments to the setup and tooling based on the platform available services.

5.10 Cost Indication

The cost of implementation is driven primarily by the usage of AWS services, including Elastic Container Service (ECS) with Fargate, Elastic Container Registry (ECR), Elastic File System (EFS) which facilitated the serverless containerized environments, and EC2 instances to host Jenkins master and Jenkins nodes. This estimate is provided based on a project that was conducted in the company (described in next section). Actual costs may vary depending on the scale of the project and especially these factors:
- Number of active branches: Each active branch has its own isolated containerized environment in ECS Fargate, which incurs cost.
- Frequency of code pushes: The frequency of code pushes to the repository led to the creation of environments, whose process incurs cost.
- Build and test process duration: Longer build and test durations meant that Fargate tasks would remain active for longer periods before being deleted, leading to higher usage costs.

As an example from one of our project proofs of concept (PoC), the monthly cost of maintaining approximately 11 branches at a time was estimated at around USD 1000. The lifespan of each branch varies, with unused or merged branches routinely deleted, so the total number of branches managed over time could be higher. This cost estimate includes expenses related to the Jenkins master and two nodes server, creation of environments for build, deployment, testing. This cost indication is based on 2024 AWS pricing and thus may vary with future changes in service costs.

6. Use Cases

The Scalable CI/CD framework has been implemented in a company project which develop a Proof of Concept (PoC) for modernization of an application which was a part of the legacy system. The PoC took form of a web application, consisting of an Angular frontend and a Java Backend for Frontend (BFF) API, and spanned a duration of six months. The team included one project manager, seven Software Engineers, two DevOps engineers, and one QA engineer. The codebase for this PoC project comprised approximately 557910 lines of code across various modules including Angular frontend components, model, and services, Java BFF Web API services, EJB client wrapper, test cases, supporting configuration and automation scripts components such as build and deployment scripts, among others, part of which was generated (boilerplates, etc.) and reused legacy codes. As this was a proof of concept, only selected functionality was implemented. We present several use cases that demonstrate the practical applications and benefits of the framework, using real examples from the project.

    The first is enhancing transparency and stakeholders engagement. In traditional development processes, particularly those involving external parties, stakeholders often lack visibility into the progress and quality of the development work. This can lead to miscommunication, delays, and an inability to verify the correctness of the delivered code. The Scalable CI/CD framework addresses this issue by providing a portal in Jenkins that tracks and displays the status of all builds and test cases in real time. Stakeholders can independently access the environment to monitor the progress of the development, including the success or failure of builds and the results of automated test cases. In this use case, the term "stakeholders" refers primarily to internal stakeholders, such as project managers, QA leads, and IT team leaders, who are directly involved in overseeing development activities. These stakeholders actively access the Jenkins portal to monitor build and test results as part of their role in the development and verification process.

    This transparency not only fosters trust between stakeholders and development teams but also empowers stakeholders to actively participate in the verification process. For instance, in cases where errors or failed test cases are identified, stakeholders can inquire about the issues and collaborate with developers to resolve them. Over time, this process helps to "open the black box" of the system, as stakeholders gain a deeper understanding of the underlying processes and dependencies.

    Another primary use case of the Scalable CI/CD framework is to facilitate the experimentation and verification of code updates in isolated environments. This capability is particularly valuable in scenarios where updates to libraries, middleware, or other dependencies are required. In the PoC project, the

developers were using a certain Maven version for the application. During an effort to test the application built using a newer Maven version, the Scalable CI/CD framework enabled the creation of a new branch and a corresponding build environment with a newer Maven version. The CI/CD pipeline automatically executed the build and deployment processes within this isolated environment.

Although the build and deployment process of the app completed successfully, the automated API tests revealed that the updated Maven version was incompatible with certain components of the system, proved by the test report showing the application failed all API tests. This discovery allowed the stakeholders to inquire the development team to identify and address the compatibility issues and make sure that the solution is not strictly version dependent, before finalizing the development. This make sure that the application can be maintained and keep on being updated in the future, hence reducing risk associated with dependency updates and future maintenance cost.

Another use case involves the migration of technologies or platforms, where the Scalable CI/CD framework provides a safe and controlled environment for testing and validation. For example, during the development of the PoC, the business required the use of JBoss as the application server. However, initial tests revealed that the JBoss server could not establish a connection with the legacy WebLogic backend. Using the Scalable CI/CD framework, we could create multiple branches to experiment with different Java versions for the JBoss server.

The experiments confirmed that connectivity could only be achieved by downgrading to a lower Java version, but this introduced runtime errors due to unsupported code functions developed with higher Java version. These findings were presented to stakeholders, who subsequently approved the decision to switch to an alternative application server that met both connectivity and functionality requirements, which has been tested and verified in another branch. This use case presents the framework's ability to support evidence-based data-driven decision-making by providing a environment for experimentation and validation.

## 7. Conclusion

The 2025 Japan Cliff poses a significant threat to Japan's economic and technological landscape, hence is a major problem that needs to be addressed. In this paper, we have developed a Scalable CI/CD Pipeline to address the internal challenges within our organization that reflect the issues identified in the 2025 Cliff problem. Our pipeline provides a solution to several issues, including:

- Automating build, deployment, and testing processes to reduce reliance on manual processes, reducing human errors and inefficiencies.
- Providing isolated and scalable development environments for each branches, enabling developers to experiment and test changes safely without risking disrupting the production system.
- Opening the black box of systems through transparency in build, deploy, test process, empowering stakeholders to verify progress and make informed decisions.

By implementing this pipeline, our organization is taking a step toward breaking free from the constraints of manual workflows and limited QA environment, which not only reduces the risks and costs associated with legacy system maintenance but also remove the psychological barriers that discourage developers from doing updates and experimentation. This marks the start of modernization toward a more sustainable software development process, which ensure that systems remain up-to-date, maintainable, and always adaptable to new emerging solution and technologies in the future.

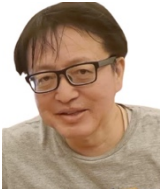

**Kuniaki Kudo**　is a Corporate Advisor at Asahi Group Holdings, Ltd.
Email:
kuniaki.kudo@asahigroup-holdings.com

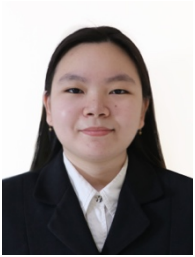

**Sherine Devi**　is an Application Engineer at Asahi Group Holdings, Ltd.
Email:
sherine.devi@asahigroup-holdings.com